# Experimental observation of coupled valley and spin Hall effect in p-doped $WSe_2$ devices

Terry Y.T. Hung[1,2], Avinash Rustagi[1], Shengjiao Zhang[1,2], Pramey Upadhyaya[1], and Zhihong Chen[1,2]*

[1]School of Electrical and Computer Engineering &

[2]Birck Nanotechnology Center

Purdue University, West Lafayette, IN 47907

*Correspondence to: zhchen@purdue.edu

Giant spin Hall effect (GSHE) has been observed in heavy metal materials such as Ta, Pt, and W, where spins are polarized in the surface plane and perpendicular to the charge current direction[1–3]. Spins generated in these materials have successfully switched magnets with in-plane magnetic anisotropy (IMA) and perpendicular magnetic anisotropy (PMA) through spin orbit torque (SOT) mechanism. It is generally accepted that PMA magnets are preferred over IMA magnets in data storage applications owing to their large thermal stability even at ultra-scaled dimensions[4]. However, SOT switching of PMA magnets by conventional GSHE materials requires either a small external magnetic field[5,6], a local dipolar field[7], or introducing tilted anisotropy to break the symmetry with respect to the magnetization[8,9]. To deterministically switch a PMA without any additional assistance, nonconventional GSHE materials that can generate spins with polarization perpendicular to the surfaces are needed. Several monolayer transition metal dichalcogenides (TMDs) have been predicted to generate such out-of-plane spins due to their 2D nature and unique band structures[10–12]. Interestingly, opposite spins are locked to their respective sub-band in each K valley of the TMD valence band with substantially large energy splitting, which enables polarized spins to be accessible through electrical gating and spatially separated by electric field through the valley Hall effect (VHE)[13–18]. Therefore, spatial separation and accumulation of spins in these 2D TMDs are uniquely referred to as coupled valley and spin Hall effect[10]. Here, we report an experiment of electrical generation of spin current with out-of-plane polarization in monolayer $WSe_2$ and detection of spin signals through a non-local spin valve structure built on a lateral graphene spin diffusion channel that partially overlaps with $WSe_2$.

Several optical measurements such as pump-probe using circularly polarized light[19–22] and Kerr rotation microscopy[23] have shown ultra-long valley life time in TMD and their heterostructures and associate that to long spin life time based on the spin-locking theory[10,11]. However, in these measurements, the spin and valley degrees of freedom are convoluted, thus do not directly probe the spin polarization. To date, only Luo *et al.*[24] succeeded in directly detecting spins by employing a MoS$_2$/graphene hybrid structure. In their experiment, valleys/spins were optically excited in monolayer MoS$_2$ with circularly polarized light and subsequently injected to a graphene channel where spins were identified through Hanle measurements. However, a direct proof of electrical generation of spins is yet to be demonstrated. It is important to note that giant spin splitting only occurs in the valence bands of few semiconducting TMDs compared to the negligible splitting in their conduction bands. In particular, WSe$_2$ is predicted to have the largest splitting of ~450meV[11]. Here, we demonstrate all electrical spin generation and detection in a WSe$_2$/graphene hybrid device and provide experimental evidences for the first time that generated spins are indeed out-of-plane polarized and locked to respective valleys that can be spatially separated. A p-type doping scheme is employed to allow easy access to sufficient amount of hole carriers in the valence band of WSe$_2$. Holes with out-of-plane spin polarization selected by the polarity of the charge current through VHE are injected to the graphene channel and finally detected by a non-local ferromagnetic contact. This is the first demonstration of an all electrical device that can generate and accumulate out-of-plane spins, which can be an important spin source for PMA based SOT-random access memory (SOT-RAM) and can possibly lead to new spin-valleytronics and novel quantum device applications.

The manipulation of entangled charge, spin, and valley degrees of freedom in TMD materials has attracted a lot of attention. Valley Hall effect has been theoretically predicted in monolayer TMDs[10,11] and experimentally demonstrated in all electrical devices by our group[17] and Wu *et al.*[18] in MoS$_2$. Broken inversion symmetry induced non-zero Berry curvature ($\Omega$) in such materials gives rise to an anomalous velocity described by $\boldsymbol{v} = \frac{e}{\hbar}\boldsymbol{E} \times \boldsymbol{\Omega}(\boldsymbol{k})$, where $\boldsymbol{k}$ is the wave vector, $\boldsymbol{E}$ is the electric field, e is the element charge, and $\hbar$ is the reduced Planck's constant. Preserved time reversal symmetry enforces opposite signs of Berry curvature in the adjacent $K$ valleys, i.e. $\boldsymbol{\Omega}(K) = -\boldsymbol{\Omega}(-K)$. A finite valley current, therefore, occurs in the transverse direction with respect to the applied electric field. Time reversal symmetry also enforces $s(K) = -s(-K)$, where s is the spin polarization, which in combination with the large spin splitting ($\Delta E \approx 450 meV$) in the valence band of monolayer WSe$_2$, gives rise to holes in the $K$ and $-K$ valley with opposite signs of spin polarization at the Fermi level. This phenomenon is called spin-valley locking[10]. Moreover, these spins have an out-of-plane spin polarization in TMDs[11]. Consequently, valley Hall effect is accompanied by a spin current with out-of-plane polarized spins flowing in a direction transverse to the applied electric field. Such out-of-plane polarized spin current can switch PMA magnets more efficiently[25]. Importantly, this type of SOT switching does not require external field assistance or special engineering of the PMA magnets.

Chemical vapor deposition (CVD) grown bilayer WSe$_2$ flakes were first transferred to a 90 (nm) SiO$_2$/Si substrate. Standard e-beam lithography was used to define Ti/Pd/Au (0.5nm/15nm/70nm) electrodes for transport measurements. A large area, continuous CVD graphene film was then transferred onto the sample covering all WSe$_2$ flakes underneath. A ferromagnetic (FM) contact with a capping layer (20nm Py / 3nmAl) was defined on top of a graphene only region by another e-beam lithography and metallization step. Al$_2$O$_3$ was then deposited by atomic layer deposition (ALD) only in a lithographically patterned rectangular area as an etching mask in order to form an isolated graphene spin diffusion channel that overlaps with a small corner of the WSe$_2$ flake. Lastly, the sample was exposed to gentle O$_2$ plasma to finally etch away the unwanted graphene region and p-dope WSe$_2$ at the same time. Details of the doping scheme are described in the reference paper[26]. In brief, the top layer of the bilayer WSe$_2$ gets converted to WO$_x$, which serves as a strong p-doping layer for the remaining WSe$_2$ layer underneath, allowing easy access to the hole carriers in the valence band (Characterizations are shown in Supplementary section I).

A schematic of the final device and its operation is illustrated in Fig. 1a along with the device's SEM image shown in Fig. 1b. Charge current ($I_{DC}$) is applied on the WSe$_2$ portion of the device indicated by the green arrow along the y direction. Valleys/spins are separated due to the valley Hall effect, resulting in a valley/spin current in the x direction with spin polarization pointing in the z direction. This current gets injected from the WSe$_2$ flake to the graphene layer sitting on top and diffuses towards the FM contact. On the right-hand side, spin chemical potentials are probed between the FM and the non-magnetic (NM) electrode. All measurements are conducted at room temperature. Raman spectra are measured to confirm the conversion of the bilayer to monolayer WSe$_2$ with the vanishing peak around 308 cm$^{-1}$ (circled), shown in Fig. 1c. Transfer characteristics of the WSe$_2$ device show a strong hole current branch indicating a successful p-doping from the converted WO$_x$ layer on top, in contrast to ambipolar characteristics with much lower hole current typically measured in monolayer WSe$_2$ devices (Fig. 1d compares the characteristics from a pristine monolayer and a converted monolayer after the O$_2$ plasma treatment). To detect the diffused spins with out-of-plane polarization, a PMA FM is a typically used in the non-local valve setup[27–29]. However, high quality PMA stacks are normally deposited by sputtering that unavoidably damages the underlying graphene due to the highly energetic sputtered particles. Alternatively, we choose e-beam evaporated Py as the FM contact, which does not damage the underlying graphene. However, this Py FM contact has in-plane magnetic anisotropy. By scanning an out-of-plane magnetic field along the z direction (B$_z$), the magnetization of the Py spin probe can be gradually pulled out-of-plane, which allows us to probe the chemical potential of the incoming out-of-plane spins. As a result, the non-local voltage changes in a continuous way when the spin probe is operated in the hard-axis, corresponding to the magnetization component along the z direction (m$_z$).

To characterize m$_z$ dependence on the applied B$_z$ field, we perform an independent anomalous Hall effect (AHE) measurement on a Py Hall bar device that has gone through the exact same fabrication processes as the test device depicted in Fig. 1a. In Fig. 2, the anomalous Hall resistance, R$_{AHE}$, increases with B$_z$ in both field directions since the magnetization of Py continuously rotates from y (in-plane, easy axis) to z (out-of-plane) direction, indicated by the arrows in the side view of the schematic AHE setup. Clear saturation is observed at large B$_z$ fields (marked as blue regions) when the magnetization of the Py electrode is completely pulled out of plane and m$_z$ reaches its maximum value, i.e. $m_z^{norm} = m_z/|max(m_z)| = \pm 1$. Knowing how the Py probe behaves under the B$_z$ field, we are now able to understand how it probes the chemical potential of the diffused spins in the graphene channel. In the device shown in Fig. 1a, b, when the DC charge current (I$_{DC}$) is applied across electrodes 1 and 2, spin current generated in WSe$_2$ through VSHE gets injected into graphene and spins diffuse towards +x direction with chemical potential of the spins exponentially decaying along the graphene channel, as shown in Fig. 3 schematic. Therefore, with the field dependent m$_z$ change in the Py FM, one can observe the change of non-local voltage (V$_{nl}$) across electrodes 3 and 4 as a function of the applied $B_z$ field. Non-local resistance, $R_{nl}$, defined by $V_{nl}$ / $I_{DC}$ as a function of applied $B_z$ field, has been measured over few devices (Supplementary Information, section II) and the average behaviors of three data sets are shown in Fig. 3a. Averaging technique is applied to enhance the signal to noise. The blue regions show clear saturation in $R_{nl}$ that can be correlated to the saturation of $m_z$ in the FM probe mentioned above. In fact, the overall behavior of $R_{nl}$ as a function of $B_z$ (green dots) can be well described by the change of $m_z$ component in the FM (black line) that is measured independently through the AHE shown in Fig. 2. Cartoons presented in Fig. 3a illustrate probing of the spin chemical potential under the FM probe when $m_z$ reaches its maximum values ($m_z^{norm} = \pm 1$) at $|B_z| > 0.8T$. When $B_z = 0T$, no non-local signals are detected since the magnetization of the FM is perpendicular to the spin polarization in the channel. With the increasing $|B_z|$, $R_{nl}$ continuously increases and reaches its maximum (minimum) value when the polarization of the spin current is parallel (anti-parallel) to the FM probe (blue regions marked in Fig. 3a). In addition, we fabricated another device with the identical structure except that the Py FM (electrode 3) was replaced with a non-magnetic electrode (NM) as our control sample. In this case, no spin chemical potential can be detected as expected, shown in Fig. 3b. The apparent differences between the test device and the control unambiguously point out that the observed increasing $R_{nl}$ and its saturation at $|B_z| > 0.8T$ is the result of the interaction between the out-of-plane spins generated in WSe$_2$ through VSHE and the FM.

To study the spin polarization under opposite electric field directions, we plot the non-local voltage reading, $V_{nl}$, as a function of $B_z$ with opposite charge current ($I_{DC}$) polarities. Anomalous velocity ($\boldsymbol{v} = \frac{e}{\hbar} \boldsymbol{E} \times \boldsymbol{\Omega}(\boldsymbol{k})$) with a given Berry curvature will change its sign when the direction of the electric field is changed. As a result, spin current with the opposite polarization will now diffuse to the graphene channel, as depicted in Fig. 4. The two possible states of spin polarization ($s_z = \pm 1$) and two FM magnetization directions ($m_z > 0$ or $< 0$) give rise to four possible combinations, labeled in their respective quadrant. To summarize our measurements, we have experimentally demonstrated for the first time that out-of-plane spins can be generated in monolayer WSe$_2$ electrically through spin-locked VHE in the valence band. For a given current polarity, the measured saturation states showing opposite $R_{nl}$ values at large $B_z$ fields are attributed to the change of the FM magnetization. On the other hand, the measured two $R_{nl}$ states under the same magnetic field with opposite current polarities are attributed to the change of the out-of-plane spin polarization in the z direction.

Our experimental findings are further verified by modeling and simulations. Given a fixed direction of DC current and a fixed direction of saturated magnetization of the FM contact, the sign of the measured non-local voltage is consistent with what to be expected from semiclassical equations of motion [Supplementary section III and IV]. Moreover, since the magnetization of the FM contact is determined by the strength of the applied perpendicular field relative to the demagnetization field, the measured non-local voltage increases in magnitude with the strength of the applied field until the magnet saturates. To determine the magnitude of the measurable voltage, we solve the drift-diffusion equation accounting for the anomalous Hall terms due to finite Berry curvature in $WSe_2$ [Supplementary section V, VI, and VII]. To get an estimate of the upper bound for the non-local voltage, we consider an ideal case where there are no losses at the $WSe_2$-graphene interface. The estimated value in this case is about ~3 times larger than the experimentally measured voltage. The difference can be attributed to the valley memory loss caused by the intervalley and valley conserving scattering at the $WSe_2$/graphene interface, leading to an interfacial spin polarization of ~ 38%.

In summary, our experimental measurements and theoretical modeling provide unambiguous evidences that valley locked out-of-plane spins can be electrically generated and accumulated, which sheds light on a possible route to achieve PMA based SOT devices or other novel all-2D spintronic devices that are currently not existing.

**Acknowledgements**

We would like to acknowledge fruitful discussions with Prof. Joerg Appenzeller and Dr. Kerem Y. Camsari. T. H., S. Z., and Z. C. gratefully acknowledge the support of this work by NEW LIMITS, a center in nCORE, a Semiconductor Research Corporation (SRC) program sponsored by NIST through award number 70NANB17H041.

**Author contributions**

Z.C. conceived and managed the research project. T.H. designed, fabricated samples and carried out electrical measurements. T.H., A.R., P.U., and Z.C. performed theoretical analysis. S.Z. carried out material preparation, optical characterizations, and partial electrical measurements. All authors discussed the results and wrote the manuscript.

**Competing interests**

The authors declare that they have no competing interests.

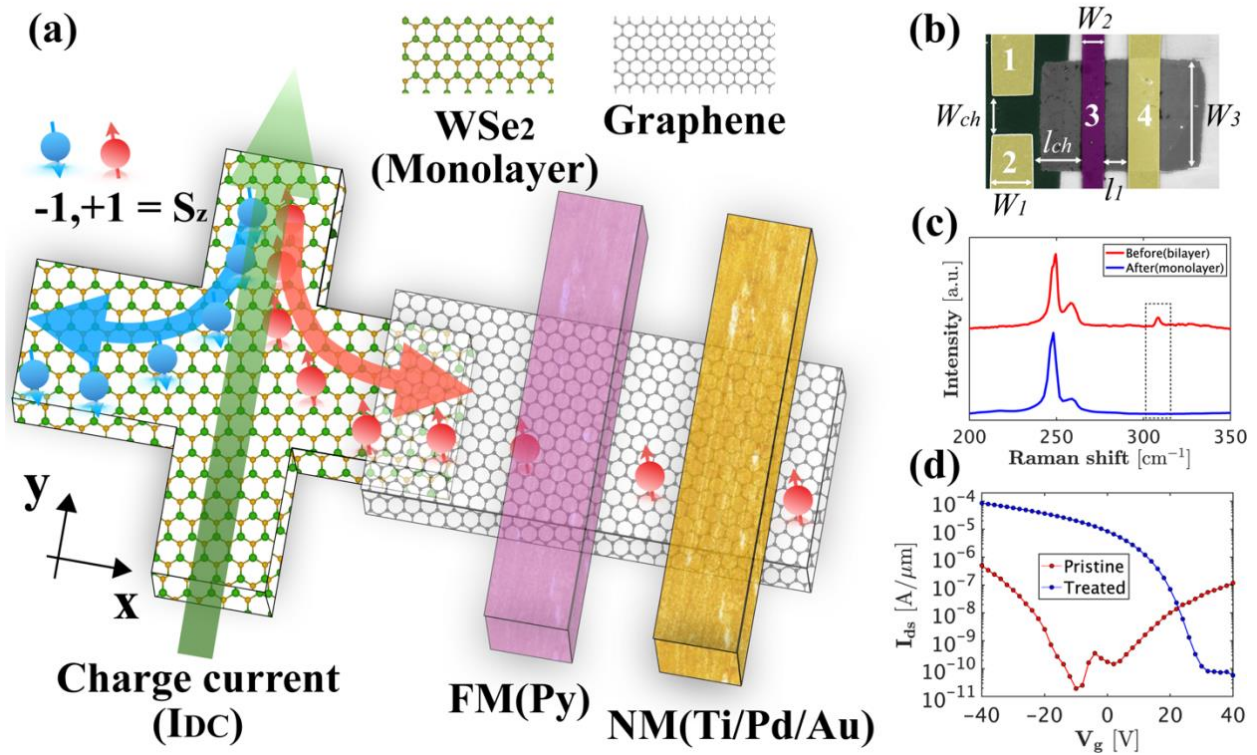

**Figure 1 | Geometry of the devices and characterizations. a,** Schematic of spin-locked valley Hall effect in monolayer $WSe_2$ and the non-local valve measurement scheme across FM and NM. **b,** Colored SEM image with detailed geometry of the devices where yellow electrodes (1, 2, 4) are NM, purple electrode (3) is FM, green area is monolayer $WSe_2$, and grey area is graphene. **c,** Raman spectrum measured on green area shows no peak ~ 308 $cm^{-1}$ (circled) which is responsible for the interlayer interaction. i.e. The peak around 308 $cm^{-1}$ will only show up when it is not monolayer. **d,** Typical transfer characteristic of pristine monolayer $WSe_2$ (red). The treated monolayer $WSe_2$ transfer characteristic measured across electrode 1 and 2 (blue).

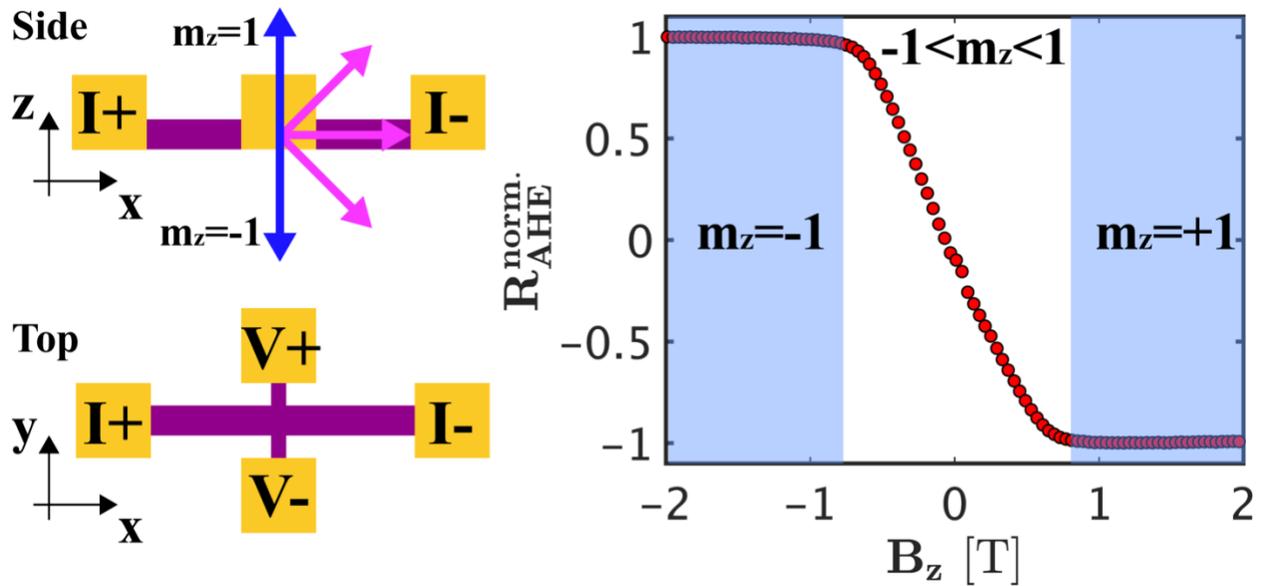

**Figure 2 | Ferromagnet characterization through anomalous Hall effect (AHE).** On the left are the side view and top view of the device where purple area is FM(Py) and yellow areas are contacting NM. Blue and magenta arrows indicate the magnetization of the FM(Py). On the right is the measured AHE which has two blue regions and one white region corresponding to the saturation of $m_z$ and the transition in between respectively.

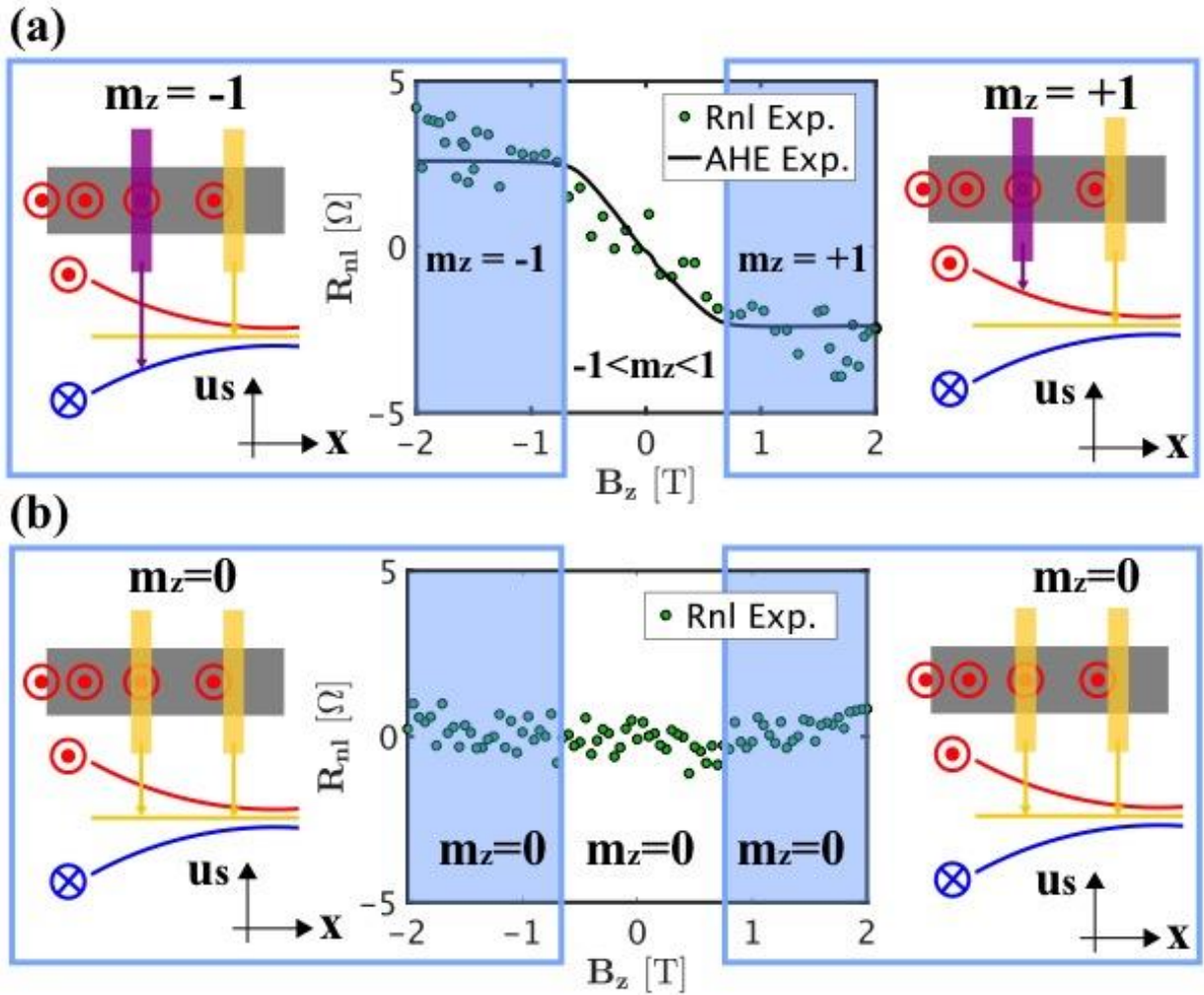

**Figure 3 | Non-local valve measurements and corresponding probing schemes. a,** Non-local measurement of the device depicted in Fig. 1a and the corresponding probing schemes at fixed spin current (polarity). Black line in the plot (named AHE Exp.) is the measurement from Fig. 2 with rescaled $R_{AHE}$ value. Red dots and blue crosses are spin polarization along +z and -z directions. Red, blue, and yellow lines are spin chemical potential of ($m_z = +1$), ($m_z = -1$), and ($m_z = 0$) respectively. **b,** Same measurement with FM electrode (purple) being replaced with NM electrode (yellow).

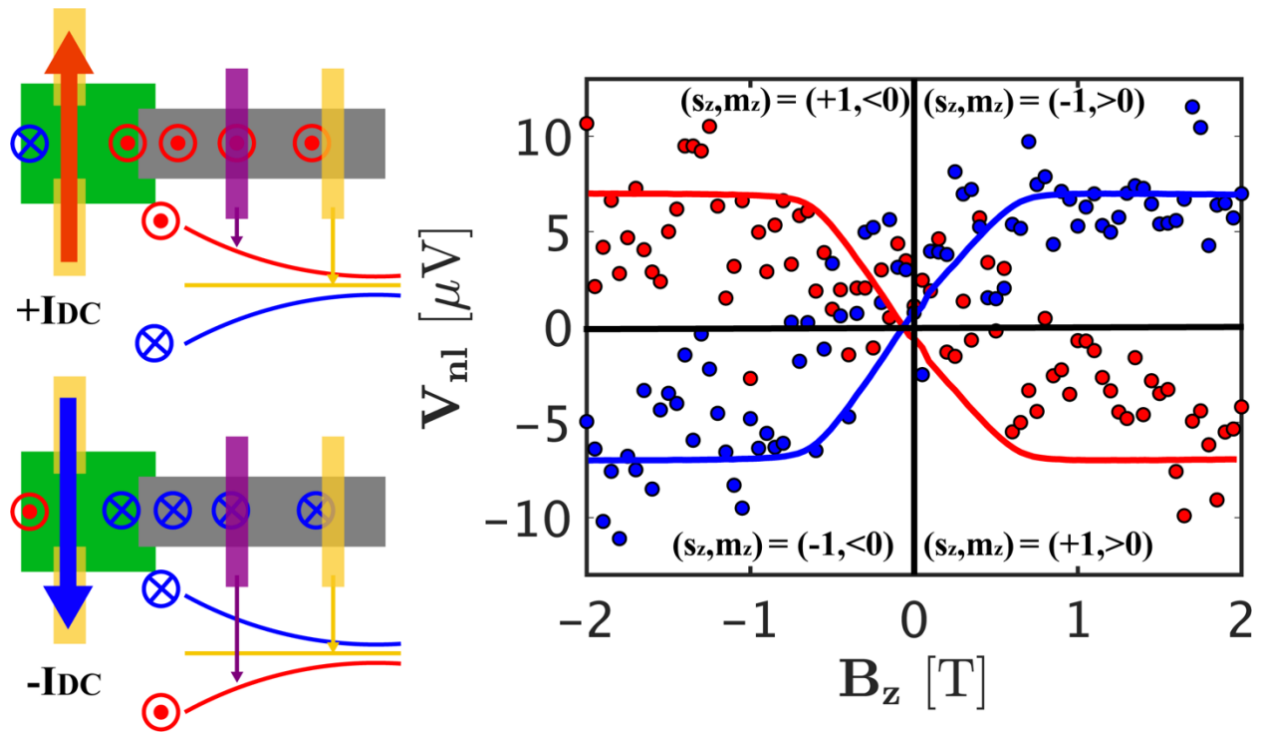

**Figure 4 | Current and spin polarity relation study through non-local valve measurements.** On the left shows the schematics of how spin current (polarity) changes with respect to the change of current polarity. On the right is the $V_{nl}$ measurements with respect to the applied $B_z$ field at opposite current polarities on the device with $l_{ch}$=1.2um, $W_1$=2um, $W_2$=1um, $W_3$=5um, and $W_{ch}$=2um. Corresponding spin polarity on the right-hand side of current path and FM magnetization ($s_z$,$m_z$) are shown in the plot. $l_{ch}$=1.2um, w1

# Supplementary of Experimental observation of coupled valley and spin Hall effect in p-doped WSe$_2$ devices

Terry Y.T. Hung[1,2], Avinash Rustagi[1], Shengjiao Zhang[1,2], Pramey Upadhyaya[1], and Zhihong Chen[1,2]*

[1]School of Electrical and Computer Engineering &

[2]Birck Nanotechnology Center

Purdue University, West Lafayette, IN 47907

*Correspondence to: zhchen@purdue.edu

## Supplementary Sections

# I. Additional Electrical and Optical Characterizations

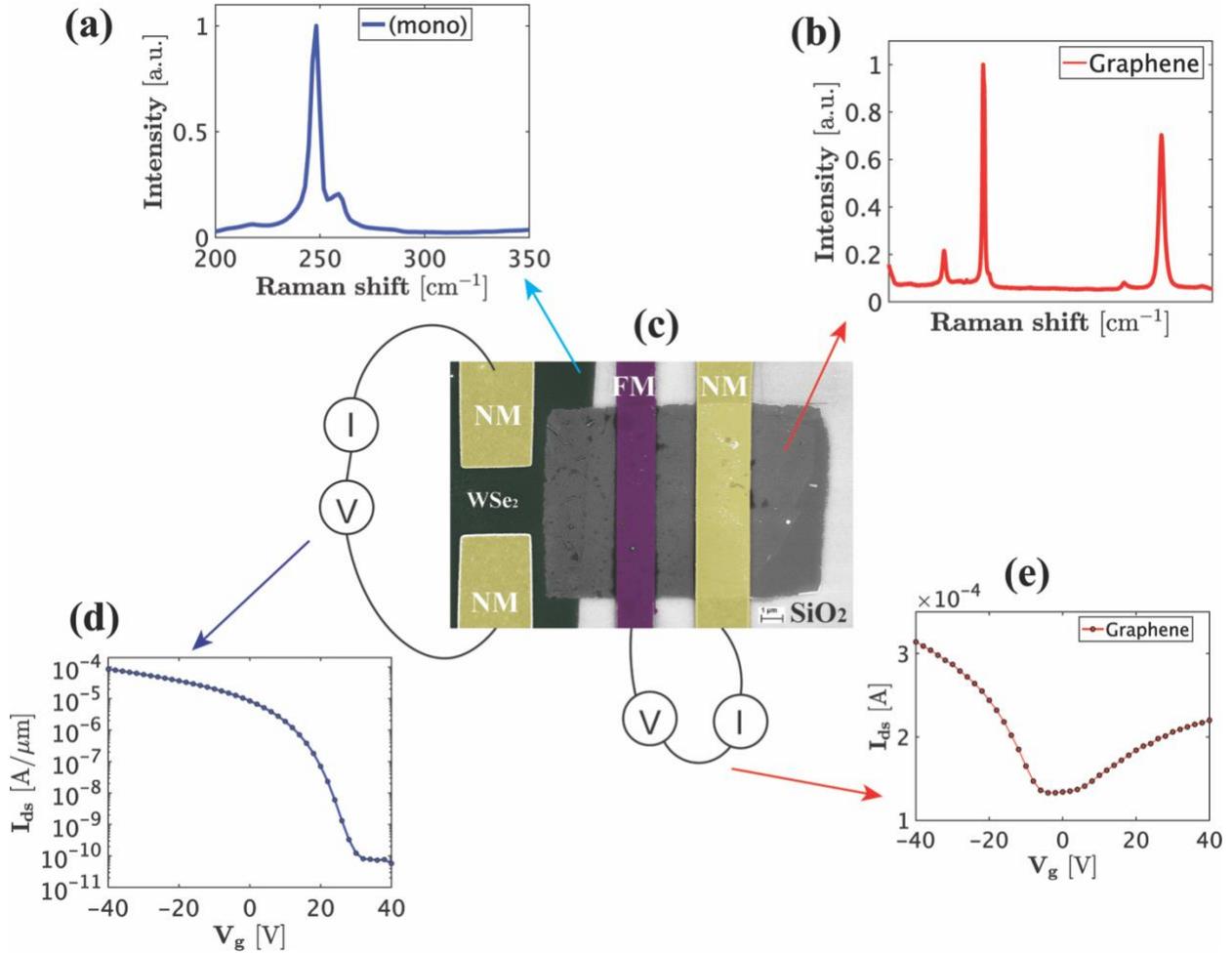

**Figure S8. | Material and Device Characterizations.** (**a, b**) Raman spectra of monolayer $WSe_2$ and graphene, respectively. (c) Colored SEM image. Yellow electrodes are non-magnetic metal electrodes of Ti/Pd/Au (0.5nm/15nm/70nm), purple electrode is the ferromagnetic contact of Py (20nm), green is the monolayer $WSe_2$, rectangle grey area is graphene, and the substrate is 90nm $SiO_2$/Si. The scale bar is 1um. (**d, e**) Transfer characteristics of monolayer $WSe_2$ and graphene, respectively.

## II. Additional Non-local Measurements on Monolayer WSe₂ Devices

Additional data sets of the same non-local measurement on monolayer WSe$_2$ devices described in the main text are shown in Fig. S7. The top three red plots are obtained when positive charge current is applied while the bottom three blue plots are for negative charge current. In Fig. S7 (d, f), linear background is subtracted.

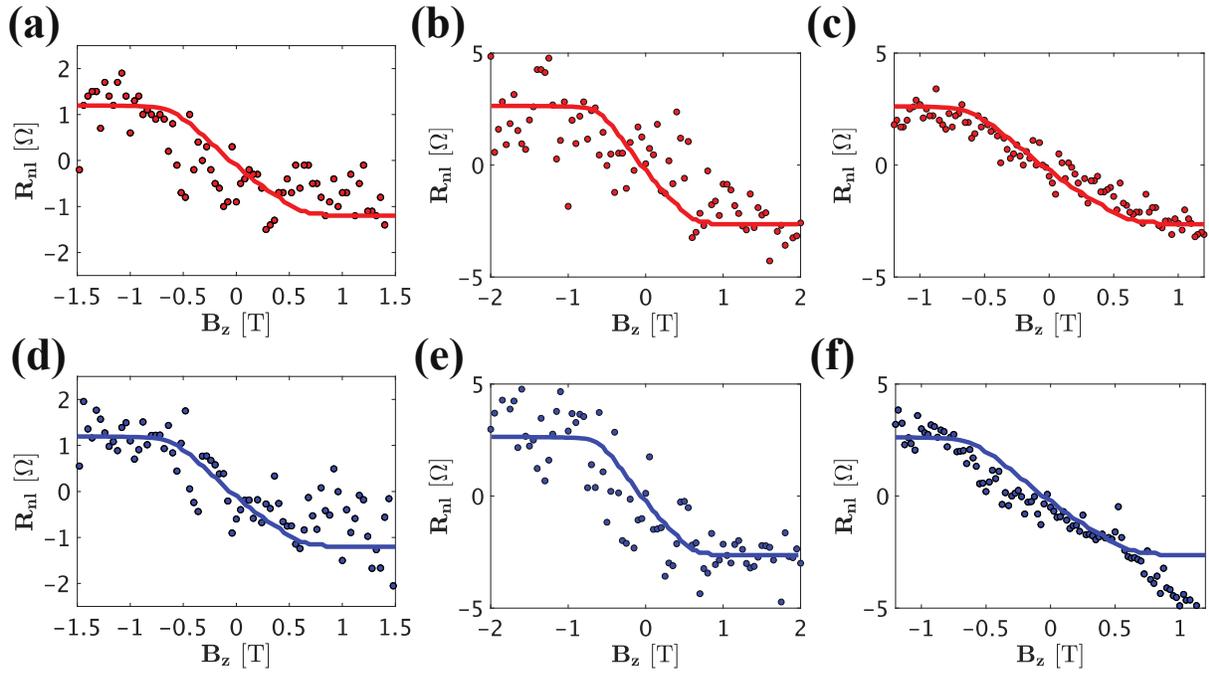

**Figure S7. | Additional Non-local Measurements.** All measurements are performed at room temperature on the monolayer WSe$_2$ devices depicted in Fig. 1b. The channel length $l_{ch}$ is 2.5, 1.2, and 1.2 um for **(a, d)**, **(b, e)**, and **(c, f)** respectively.

## III. Sign of the Measured Non-local Voltage

The finite Berry curvature of the bands allows for valley Hall effect which can be measured electrically based on the device schematic in Fig. S2. Based on the information at hand, it is possible to evaluate the sign of the non-local voltage when the magnet is completely saturated out of plane by the application of a large external field. The semiclassical equations that govern the Bloch electron wavepacket dynamics for electron in the n-th band with wavevector $\vec{k}$ and position $\vec{r}$ and experiencing real space electric and magnetic fields[1] are

$$\dot{\vec{r}} = \frac{1}{\hbar}\frac{\partial \vec{E}}{\partial \vec{k}} - \dot{\vec{k}} \times \vec{\Omega}$$

$$\hbar\dot{\vec{k}} = -e\vec{E} - e\dot{\vec{r}} \times B$$

(5)

Corresponding to the non-local measurement, a charge current is driven along the $+\hat{y}$ by application of an electric field $\vec{E} = E\hat{y}$ through p-doped WSe$_2$. Due to the large VB spin splitting in monolayer WSe$_2$, only one of the spin split bands from each inequivalent K/-K valley participate in transport. In absence of any external magnetic field, the equation of motion simplifies to

$$\dot{\vec{r}} = \frac{1}{\hbar}\frac{\partial \vec{E}}{\partial \vec{k}} + \frac{e\vec{E}}{\hbar}\hat{y} \times \vec{\Omega}_v$$

$$= \frac{1}{\hbar}\frac{\partial \vec{E}}{\partial \vec{k}} + \frac{eE}{\hbar}\Omega_v \tau \hat{x},$$

(6)

thus

$$\dot{x} \propto \frac{eE\Omega_v}{\hbar}\tau.$$

(7)

Hence, for the K-valley electrons ($\tau = +1$): $\dot{x} > 0$ and those are right moving. Due to spin-split electronic bands in the valley, we can conclude that the K−valley electrons participating in transport are of the spin up type.

There are a few key steps in understanding what the magnet measures (spin up/down electrochemical potentials).

To resolve this, let us consider the case of magnetization pointing along the +z direction. Magnetization along +z implies the electron spins in ferromagnet (FM) are along -z (reason the negative gyromagnetic ratio for electrons). Since the electron spins in magnet are along -z, the spin down electrons are able to tunnel into/from the magnet and equilibrate. The non-magnetic (NM) metal always measures the average electrochemical potential of the up and down spin electrons. Thus, the non-local voltage defined as $V_{nl} = V_{NM} - V_{FM} = \frac{\mu_{NM} - \mu_{FM}}{-e}$. Due to the established equilibrium, the ferromagnet measures the lower electrochemical potential for spin down electrons. Thus, since $\mu_{NM} < \mu_{FM}$, the non-local voltage $V_{nl}$ measured is < 0.

In analogy, when the magnetization of the magnet is along the -z direction, the non-local potential $V_{nl} = V_{FM} - V_{NM} > 0$.

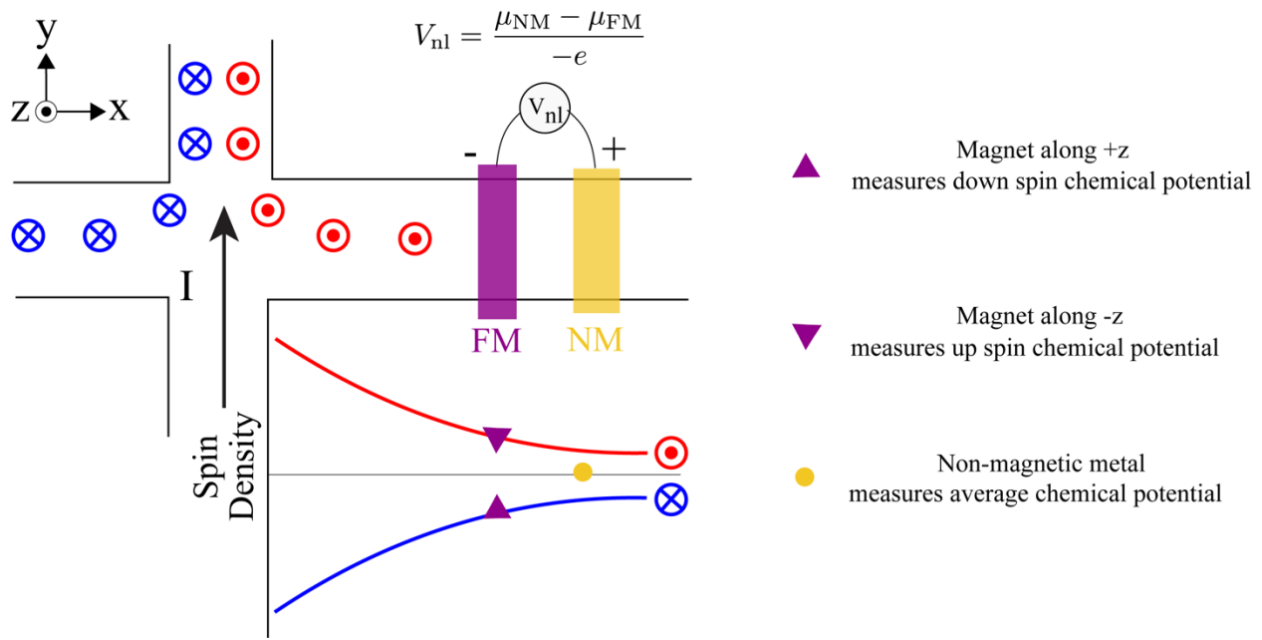

**Figure S2. | Non-Local Measurement Geometry for the Electrical Detection of coupled VSHE.**

## IV. Perpendicular Field Dependence of Non-local Voltage

Now that we have established the sign of the non-local voltage given the magnetization polarized completely out of the plane, the next step is to relate $V_{nl}$ to the applied perpendicular field. When magnetization is completely oriented in the +z direction, the non-local potential attains the maximum negative value $V_{nl} = -V_{nl}^0$ (where $V_{nl}^0 > 0$). And when oriented in the -z direction, the non-local potential attains the maximum positive value $V_{nl} = V_{nl}^0$. Thus, we can simply conclude that $V_{nl}$ as a function of magnetization orientation follows $V_{nl} = -V_{nl}^0 \cos\theta$, where θ is the polar angle of the magnetization with the out of plane z-direction. To relate the magnetization orientation given the applied perpendicular field, we can minimize the free energy within the macrospin approximation. Since permalloy is a weak magnet, the free energy describing the magnet consists of the external field term (Zeeman) and the demagnetization term accounting for the large shape anisotropy cost for out-of-plane magnetization,

$$\mathcal{F} = -M_s B_z m_z + 2\pi M_s^2 m_z^2 = -M_s B_z \cos\theta + 2\pi M_s^2 \cos^2\theta \tag{8}$$

Minimizing the free energy with respect the magnetization orientation

$$\frac{\partial \mathcal{F}}{\partial \theta} = M_s B_z \sin\theta - 4\pi M_s^2 \sin\theta \cos\theta = 0$$

$$\Rightarrow \cos\theta = \begin{cases} \dfrac{B_z}{4\pi M_s} & \text{when } |B_z| < 4\pi M_s \\ \text{sign}(B_z) & \text{when } |B_z| > 4\pi M_s. \end{cases} \tag{9}$$

Thus, the non-local voltage depends on the applied perpendicular field as shown in Fig. S3,

$$V_{nl}(B_z) = \begin{cases} -V_{nl}^0 \dfrac{B_z}{4\pi M_s} & \text{when } |B_z| < 4\pi M_s \\ -V_{nl}^0 \text{sign}(B_z) & \text{when } |B_z| > 4\pi M_s. \end{cases} \tag{10}$$

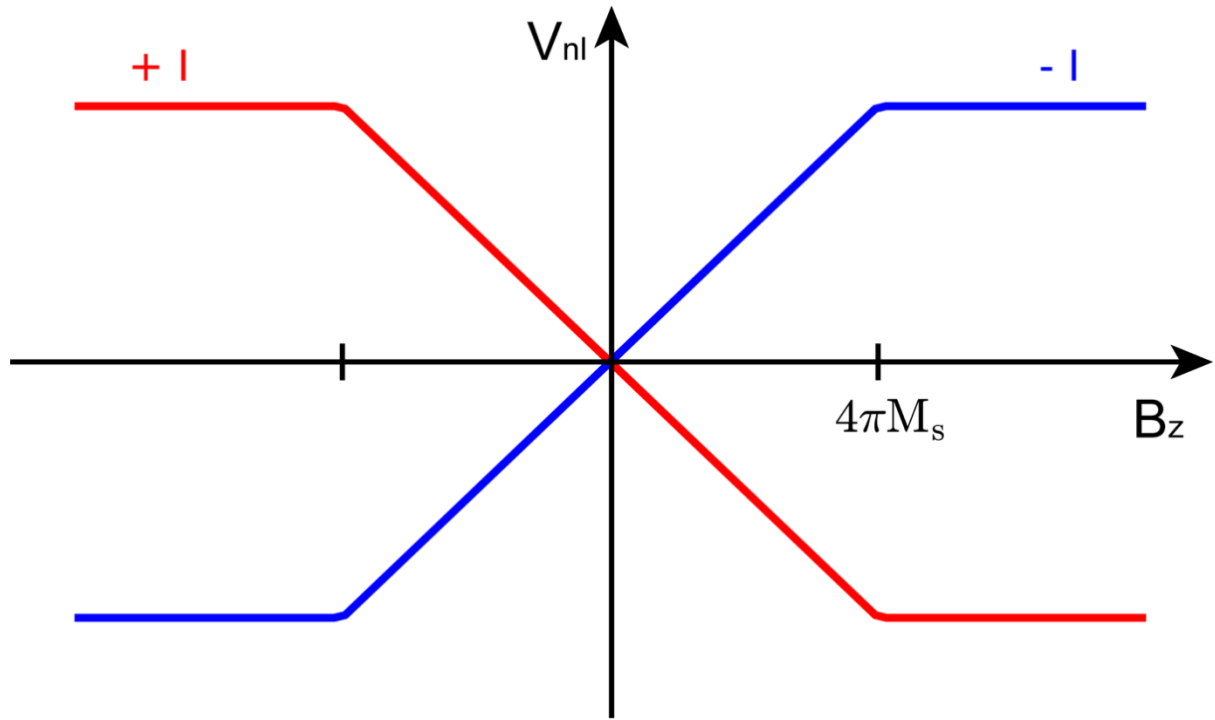

**Figure S3. | Sketch of Non-Local Voltage Measured as a Function of Perpendicular Magnetic Field with Positive Charge Current (Red) and Negative Charge Current (Blue).**

## V. Valley Hall Conductivity

The study of spin-locked valley Hall effect is characterized by the valley Hall conductivity which is defined as[2]

$$\sigma_{xy}^{VH}(\mu) = \sum_\tau \tau \sum_{s_z,\alpha} \frac{e^2}{\hbar} \int \frac{d\vec{k}}{(2\pi)^2} \vec{\Omega}(\vec{k}) f(E - \mu), \quad (11)$$

where $\mu$ is the chemical potential, $\alpha = \pm 1$ is the band index CB/VB, $s_z = \pm 1$ is the spin index up/down, and $\tau = \pm 1$ is the valley index K/K'. Considering the TMDC is p-doped, the chemical potential lies below the mid-energy gap i.e. $\mu < 0$. Thus, only the valence band at the valleys are contributing to the valley Hall conductivity. The valley Hall conductivity calculated as a function of chemical potential is shown in Fig. S4.

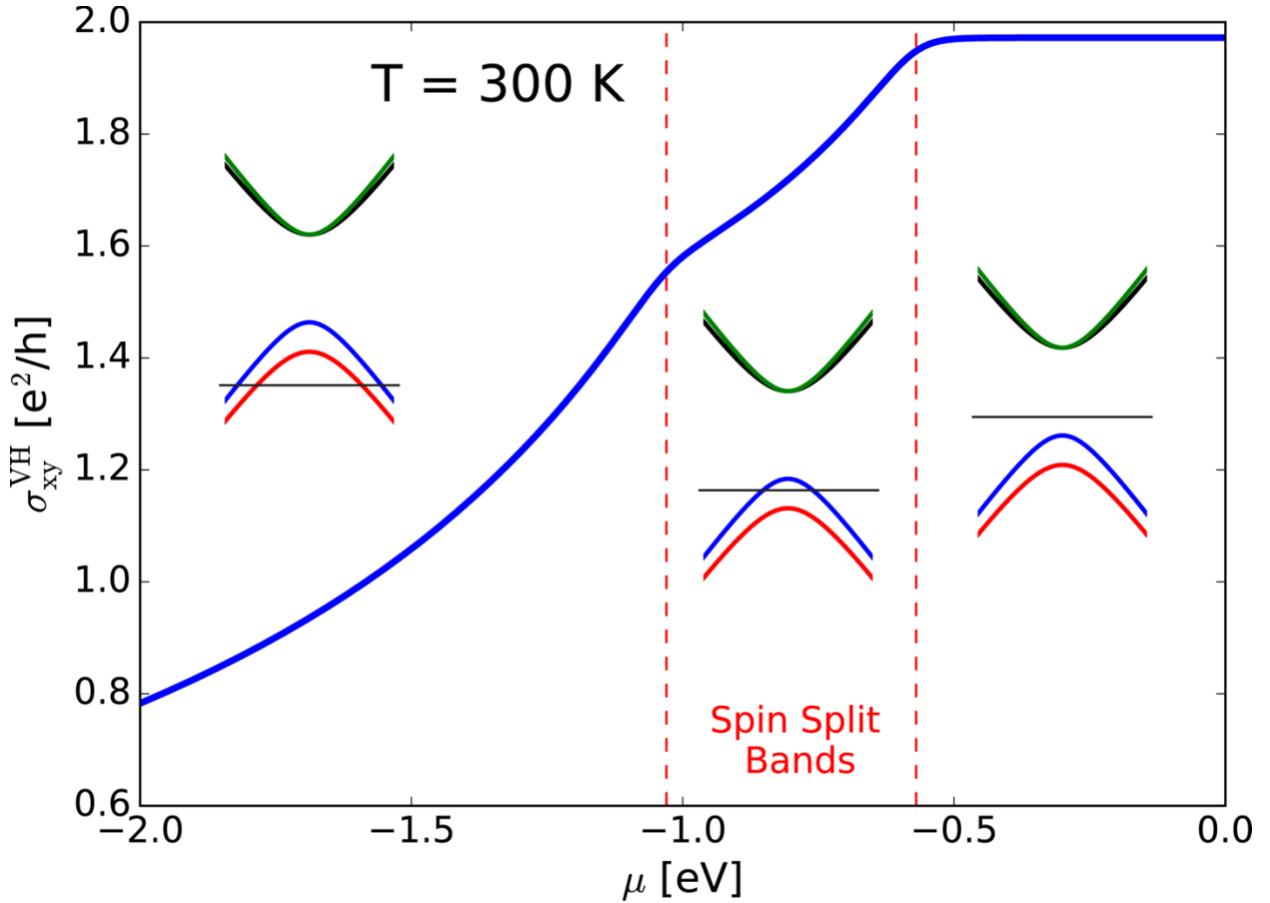

**Figure S4. | Valley Hall Conductivity Calculated as a Function of Chemical Potential.**

## VI. Theoretical Model for Calculating Non-local Voltage

Our goal here is to model the non-local voltage measured by the FM/NM electrodes for which we will need to estimate the spatially varying valley chemical potential difference $\delta\mu_v = \mu_K - \mu_{K'}$. Here we follow the model in [3]. The scheme here is to divide the device into regions as shown in Fig. S5. In each region, the diffusion equation describing the valley chemical potential difference $\delta\mu_v = \mu_K - \mu_{K'}$ is solved self-consistently with the conductance matrix equation that accounts for the source term. The valley chemical potential difference $\delta\mu_v = \mu_K - \mu_{K'}$ follows the diffusion equation

$$\frac{\partial^2}{\partial x^2}\delta\mu_v^i = \frac{1}{(l_v^i)^2}\delta\mu_v^i, \tag{12}$$

where region $i = \{A, B, C\}$ and $l_v^i$ is the inter-valley scattering length (processes by which the valley polarization is lost). The regions A and B are WSe$_2$ and region C is graphene. The equation relating the charge current (along y-direction) and valley current (along x-direction) to their corresponding generalized forces namely electric field (along y-direction) and gradient of valley chemical potential difference (along x-direction) are related as

$$\begin{pmatrix} j_c^i \\ j_v^i \end{pmatrix} = \begin{pmatrix} \sigma_{xx,i} & -\sigma_{xy,i}^{VH} \\ \sigma_{xy,i}^{VH} & \sigma_{xx,i} \end{pmatrix} \begin{pmatrix} E_i \\ -\frac{1}{2e}\frac{\partial}{\partial x}\delta\mu_v^i \end{pmatrix}. \tag{13}$$

Typically, when the materials are pure, the inter-valley scattering are caused by the edge of the devices. Thus, for the device geometry shown, $l_v^A = l_v$, $l_v^B = \infty$ (since there are no edges assuming contacts for charge current flow are ideal), and $l_v^C = l'_v$. The WSe$_2$ regions A and B have non-zero valley hall conductance and are characterized by $\sigma_{xx,A/B} = \sigma_{xx}$ and $\sigma_{xy,A/B}^{VH} = \sigma_{xy}^{VH}$. For region C i.e. graphene, no berry curvature implies absence of valley hall conductance. Therefore, $\sigma_{xx,C} = \sigma'_{xx}$ and $\sigma_{xy,C}^{VH} = 0$. The boundary conditions are $\delta\mu_v^A(x = -\infty) = 0$ and $\delta\mu_v^C(x = +\infty) = 0$.

As for the interfaces, the assumed conditions are continuity of valley chemical potential difference and continuity of valley current across at the interface of region A and B. However, the WSe$_2$-graphene interface in general lead to the boundary conditions of discontinuity in valley chemical potential and valley current resulting in valley memory loss. This is similar to the spin memory loss due to interfaces when studying spin hall effect[4]. Analogous to spin hall, the discontinuity in valley chemical potential difference can be thought of as the asymmetry in interface resistance corresponding to the two different valley electrons due to valley conserving scattering. This leads to potential drop across the resistances and thus corresponds to discontinuity in valley chemical potentials. On the other hand, the interface can also lead to valley flip scattering and this in turn implies discontinuity in valley current illustrated through the two-channel resistor model in Fig. S6. Thus, in the simplest of models for the interface, we can parameterize the interface effects by means of the opacity parameters α and β corresponding to discontinuity in valley chemical potential difference and valley current. The charge current and electric field in region B is $j = I/w$ and $E$ respectively. The interface conditions thus imply,

$$\delta\mu_v^A(x = -w/2) = \delta\mu_v^B(x = -w/2)$$
$$(1-\alpha)\delta\mu_v^B(x = w/2) = \delta\mu_v^C(x = w/2),$$
(14)

where 1−α corresponds to the transparency of the valley chemical potential across the WSe$_2$-graphene interface, and for the valley current

$$j_v^A(x = -w/2) = j_v^B(x = -w/2)$$
$$(1-\beta)j_v^B(x = w/2) = j_v^C(x = w/2),$$
(15)

where 1 − β corresponds to the transparency of the valley current across the WSe$_2$-graphene interface. The diffusion equation is solved in a self-consistent manner alongside the conductance equation (which accounts for the source term) to determine the spatially varying valley chemical potential difference in graphene that is measured,

$$\delta\mu_v^C = I\exp\left(-\frac{x-w/2}{l'_v}\right)\frac{2el'_v\sigma_{xy}^{VH}}{\sigma_{xx}\sigma'_{xx}}\frac{1-\beta}{\left(\frac{1-\beta}{1-\alpha}\right)\frac{l'_v}{l'_v}\left[\frac{(\sigma_{xy}^{VH})^2}{\sigma_{xx}}+\sigma_{xx}\right]+l_v+w}. \tag{16}$$

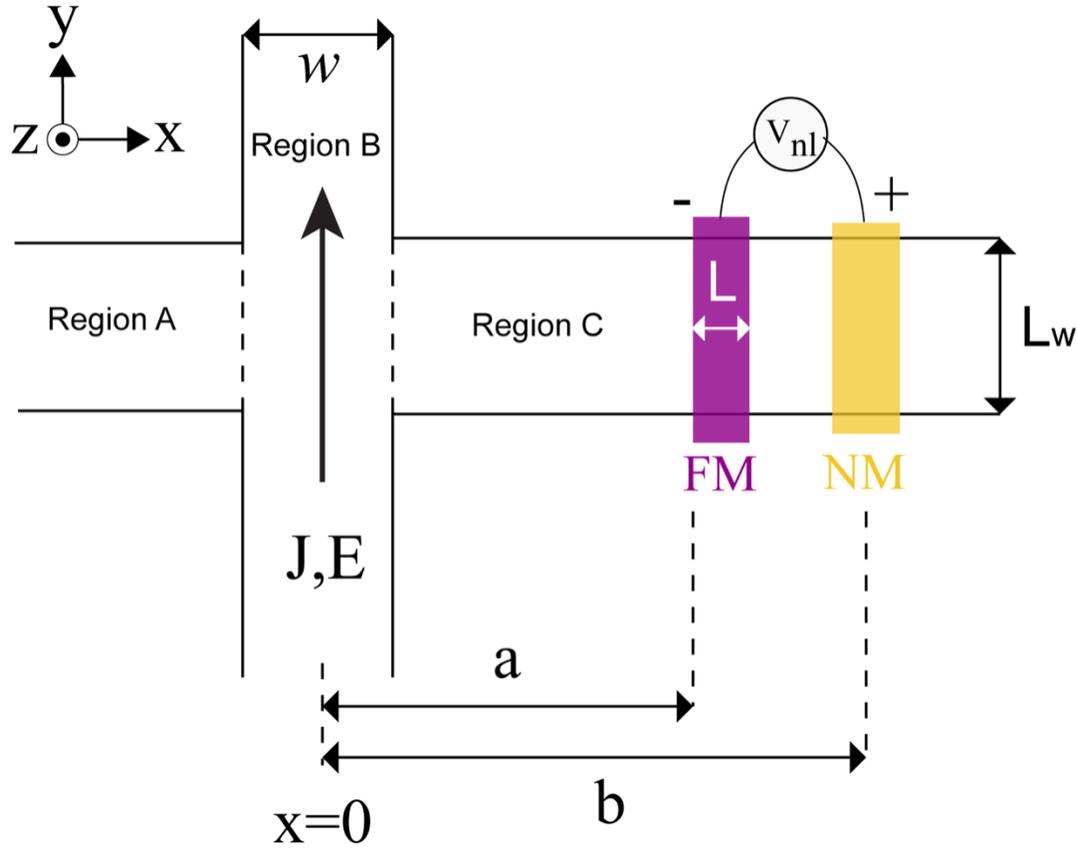

**Figure S5. | Device Schematic for Non-local Electrical Generation and Detection of VSHE.**

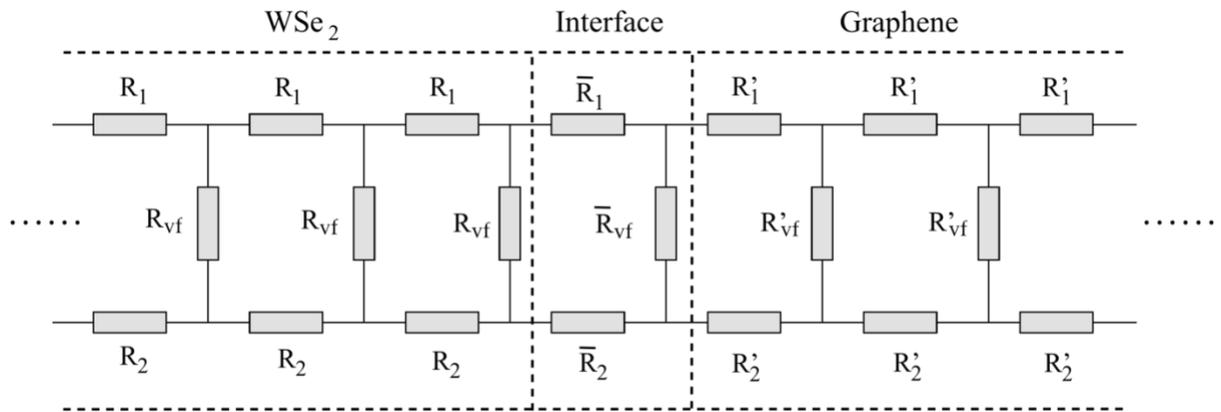

**Figure S6. | Resistor Two-channel Model for the WSe$_2$/Graphene Interface, Accounting for the Discontinuity in Chemical Potential ($\bar{R}_1, \bar{R}_2$) and Valley Current ($\bar{R}_{vf}$).**

## VII. Numerical Estimate: Upper Bond

To make an estimate for the upper bound for the saturated non-local voltage amplitude, we consider the most pristine case where there are no losses at the interface (i.e. $\alpha = \beta = 0$). WSe2 and graphene have similar spin diffusion lengths (i.e. $l_v = 0.6$ μm and $l_{v'} = 1$ μm) and longitudinal conductivities $\sigma_{xx} = 2.5$ e² / h ($\equiv 10k\Omega$) and $\sigma'_{xx} = 5$ e² / h ($\equiv 5k\Omega$). For the Valley Hall $\sigma_{xy}^{VH} \approx 2e^2/h$. The DC current used in measurement is 2 μA. The left edge of the ferromagnet electrode is at a=1.2μm and of width L=1μm. Thus, the average chemical potential it measures is

$$\langle \delta\mu_v^C \rangle = \frac{1}{L}\int_{a+\frac{w}{2}}^{a+\frac{w}{2}+L} dx \delta\mu_v^C(x) = \frac{1}{L}\frac{2el'_v{}^2\sigma_{xy}^{VH}}{\sigma_{xx}\sigma'_{xx}} \frac{e^{-a/l'_v}-e^{-(a+L)/l'_v}}{\frac{l'_v}{\sigma'_{xx}}\left[\frac{\left(\sigma_{xy}^{VH}\right)^2}{\sigma_{xx}}+\sigma_{xx}\right]+l_v+w}. \quad (17)$$

Assuming the contact width w = 2um, and that no charge current flowing along the valley current in graphene, there is a symmetry in the chemical potential of the K and -K valleys. Therefore $\langle \delta\mu_v^{\vec{K}} \rangle = \langle \delta\mu_v^C \rangle/2$ and the non-local voltage is

$$V_{nl}^0 = \frac{\langle \delta\mu_v^C \rangle}{2e} = 4.6 \times 10^{-4} \, V. \quad (18)$$

Here, to determine the upper bound of non-local voltage, we have assumed that the magnet is able to sense the chemical potential with unit efficiency. This is true if one of the spins sub-bands in the ferromagnet is full and there is no spin relaxation at the graphene-ferromagnet interface. However, in a general setting, if none of the spins sub-bands in the ferromagnet is full, the measured non-local voltage is reduced by the polarization efficiency of the detector ferromagnet $P_D$[5] which was determined from experiments on graphene non-local spin valves in [6] to be ~ 4%. Thus the upper bound on the measured non-local voltage is

$$V_{nl}^m = P_D V_{nl}^0 = P_D \frac{\langle \delta\mu_v^C \rangle}{2e} = 1.84 \times 10^{-5} \, V. \quad (19)$$

To summarize, the difference between the actual measured non-local voltage and the theoretical upper bound estimation can be attribute to the quality of the WSe$_2$-graphene interface.